\documentclass[twocolumn,showpacs,preprintnumbers,amsmath,amssymb,pra]{revtex4}
\usepackage{graphicx}
\usepackage{dcolumn}
\usepackage{bm}

\begin{document}


\title{Experimental demonstration of quantum teleportation of a squeezed state}

\author
{Nobuyuki Takei,$^{1,2}$ Takao Aoki,$^{1,2}$ Satohi Koike,$^{1}$ Kenichiro Yoshino,$^{1}$ Kentaro Wakui,$^{1}$ Hidehiro Yonezawa,$^{1,2}$ Takuji Hiraoka,$^{1}$ Jun Mizuno,$^{2,3}$ Masahiro Takeoka,$^{2,3}$ Masashi Ban,$^{2,4}$ and Akira Furusawa$^{1,2}$}

\affiliation{
$^{1}$Department of Applied Physics, School of Engineering, The University of Tokyo,\\
7-3-1 Hongo, Bunkyo-ku, Tokyo 113-8656, Japan\\
$^{2}$CREST, Japan Science and Technology (JST) Agency, 1-9-9 Yaesu, Chuo-ku, Tokyo 103-0028, Japan\\
$^{3}$National Institute of Information and Communication Technology (NICT), 4-2-1 Nukui-Kitamachi, Koganei, Tokyo 184-8795, Japan\\
$^{4}$Advanced Research Laboratory, Hitachi Ltd, 2520 Akanuma, Hatoyama, Saitama 350-0395, Japan
}

\date{\today}

\begin{abstract}
Quantum teleportation of a squeezed state is demonstrated experimentally. 
Due to some inevitable losses in experiments, a squeezed vacuum necessarily becomes a mixed state which is no longer a minimum uncertainty state. 
We establish an operational method of evaluation for quantum teleportation of such a state using fidelity, and discuss the classical limit for the state. 
The measured fidelity for the input state is 0.85$\pm$ 0.05 which is higher than the classical case of 0.73$\pm$0.04. 
We also verify that the teleportation process operates properly for the nonclassical state input and its squeezed variance is certainly transferred through the process. 
We observe the smaller variance of the teleported squeezed state than that for the vacuum state input. 
\end{abstract}


\maketitle

\section{INTRODUCTION}

Quantum teleportation enables reliable transfer of an unknown quantum state from one location to another~\cite{Bennett93}. 
This transfer is achieved by utilizing shared quantum entanglement and classical communication between two stations. 
The initial approaches using qubits~\cite{Bennett93,bouwme97} have been extended to a continuous-variable (CV) system~\cite{vaidman94,braun98} employing the Einstein-Podolsky-Rosen (EPR) correlation~\cite{einstein35}. 
So far several experiments for CVs have been demonstrated for a coherent state input using quadrature-phase amplitudes of an electromagnetic field mode~\cite{furusawa98,bowen03a,zhang03,takei05}. 
Teleportation of quantum entanglement, i.e., entanglement swapping has been also realized~\cite{takei05,jia04}. 
Furthermore, CV teleportation has been extended to a multipartite protocol known as a quantum teleportation network~\cite{vanloock00,yonezawa04}.

The experiments with a coherent state input have been performed by assuming that the input is a pure state. 
For such a state, there has been much investigation of fidelity as a success criteria, and its value and the classical limit are well understood~\cite{braun00,braun01,Ban04,Grosshans01}. 
However the input state would be not always pure but mixed due to some inevitable losses and imperfection in real experiments. 
Furthermore there may be a situation in which a manipulated state in some imperfect quantum circuits will be teleported as an input. 
In this case, the state would be considered as a mixed state. 
But the success criterion for CV teleportation of a mixed state input has not been investigated very much so far.

In entanglement swapping, a subsystem of a bipartite entangled state is teleported. 
Since the subsystem is generally a mixed state, entanglement swapping is considered as teleportation of a mixed state. 
The success of this teleportation has been verified by examining quantum entanglement between the output and the partner subsystem~\cite{takei05,jia04} in terms of the inseparability criterion~\cite{duan00,simon00}. 
However this verification is applicable only to the case of a bipartite entangled state or a two-mode state, and is not applicable to the case of a single-mode mixed state input. 
Therefore an operational method of verification for a single-mode mixed state should be established.

The previous experiments for CVs have been carried out using Gaussian states, i.e., those states whose Wigner functions have Gaussian distribution on the phase space. 
Even when a Gaussian state suffers from some losses and becomes a mixed state, the state is just transformed into another Gaussian state. 
So we focus on only Gaussian states. 
In such states, single-mode mixed states are provided as displaced squeezed thermal states~\cite{adam}. 
The fidelity for these states has been studied~\cite{twamley,Jeong} and applied to quantum teleportation~\cite{ban04}. 
However the success criterion or the classical limit has not been understood as mentioned above.

In this paper, we experimentally demonstrate CV teleportation of a squeezed vacuum which belongs to Gaussian states. 
Due to some inevitable losses in real experiments, a squeezed vacuum is degraded and necessarily becomes a mixed state which is no longer a minimum uncertainty state and called a squeezed thermal state. 
However, as long as its squeezed variance is smaller than the vacuum variance, we call the mixed state a squeezed vacuum in the present work. 
Note that general squeezed thermal states include both the squeezed vacuum states and the states with the squeezed variances larger than the vacuum variance.

We investigate quantum teleportation of a squeezed vacuum state and calculate the fidelity given in Refs.~\cite{twamley,Jeong,ban04}. 
We establish an operational method of evaluation for the teleportation of the mixed state and discuss the classical limit for the state. 
Moreover we also verify that the teleportation process operates properly for a squeezed vacuum input, and we observe that the squeezed variance of the input is certainly teleported.

This paper is organized as follows: In Sec. II, we describe a squeezed state which is an input state in our teleportation. In Sec. III, we briefly summarize the procedure of quantum teleportation and explain in detail our experimental set-up. In Sec. IV, we show the experimental results and calculate the fidelity. In Sec. V, we discuss the classical limit for a set of mixed squeezed states. Section VI is for conclusion.

\section{A SQUEEZED THERMAL STATE}

We consider quantum teleportation of a squeezed vacuum state of an electromagnetic field mode. 
The field mode can be represented by an annihilation operator $\hat{a}$~\cite{WM}, whose real and imaginary parts ($\hat{a}=\hat{x}+i\hat{p}$) correspond to quadrature-phase amplitude operators with the canonical commutation relation $[\hat{x}, \hat{p}]=i/2$ (units-free, with $\hbar=1/2$). 
In this notation, the variances of a vacuum state are given by $\langle (\Delta \hat{x})^2 \rangle_{vac}=\langle (\Delta \hat{p})^2 \rangle_{vac}=1/4$. 
A squeezed vacuum state is defined as the state which has reduced variance from the vacuum variance in one quadrature at the expense of increased variance in the other, for example, $\langle (\Delta \hat{x})^2 \rangle < 1/4 <\langle (\Delta \hat{p})^2 \rangle$~\cite{WM}. 
This state belongs to the class of minimum-uncertainty states: $\langle (\Delta \hat{x})^2 \rangle \times \langle (\Delta \hat{p})^2 \rangle =1/16$.

However, for any real experiments, one usually obtains a squeezed vacuum state which suffers from some inevitable losses. 
Such a squeezed vacuum is not a pure but mixed state. 
This is certainly true of our squeezed vacuum input. 
This mixed squeezed vacuum is regarded as a squeezed thermal state. 
Assuming that $x$ quadrature is squeezed, its variances are written as follows: 
\begin{equation}
\left\{
\begin{array}{ll}
\sigma_{\mathrm{in}}^{x} = \langle (\Delta \hat{x}_{\mathrm{in}})^2 \rangle =e^{-2r}\coth (\beta/2)/4 \\
\sigma_{\mathrm{in}}^{p} = \langle (\Delta \hat{p}_{\mathrm{in}})^2\rangle = e^{+2r}\coth (\beta/2)/4
\end{array}
\right.
,
\label{eq:sq}
\end{equation}
where $r$ is the squeezing parameter and $\coth (\beta/2)/4$ is the variance of an initial thermal state. 
$\beta$ is the inverse temperature $1/2k_{{\mathrm B}}T$ where $k_{{\mathrm B}}$ is the Boltzmann constant and $T$ is temperature. 
Accordingly a squeezed thermal state is no longer the minimum-uncertainty state: $\sigma_{\mathrm{in}}^{x} \times \sigma_{\mathrm{in}}^{p} > 1/16$.

Let us consider quantum teleportation of a more general squeezed thermal state, i.e., a displaced squeezed thermal state with some rotation in the phase space. 
This state may be also the most general Gaussian state~\cite{adam}. 
The Wigner function of such a Gaussian state is given by~\cite{Jeong}
\begin{eqnarray}
W(x',p') &=&\frac{1}{2\pi \sqrt{\sigma^x_{\mathrm{in}} \sigma^p_{\mathrm{in}}}} \exp \Bigl\{ -\frac{1}{2 \sigma^x_{\mathrm{in}}}  \left( x'-x_0 \right)^2 \nonumber \\
&&{}- \frac{1}{2\sigma^p_{\mathrm{in}}} \left( p'-p_0 \right)^2 \Bigr\},
\label{eq:wigner}
\end{eqnarray}
where $x',\ p'$ are coordinates rotated from $x$ and $p$ axes by an angle $\theta$ in the phase space: $x'=x\cos \theta +p\sin \theta,\ p'=-x\sin \theta +p\cos \theta$. 
$x_0$ and $p_0$ represent displacement of the state $\alpha_0 = x_0 +i p_0$. 
From Eqs.~(\ref{eq:sq}) and (\ref{eq:wigner}), the Gaussian state can be fully characterized by four parameters $r$,\ $\beta$,\ $\theta$ and $\alpha_0$.

In a teleportation process, displacement $\alpha_0$ of an input state can be easily reconstructed at the output station by setting gains of classical channels to unity. 
This exact reconstruction of the displacement is the definition of unity gain. 
Note that the unity gains are calibrated by using a strong field with a sufficiently large displacement $|\alpha_0|\gg 1$ which is treated classically~\cite{zhang03}. 
It follows from unity gain that the fidelity, which is the overlap between input and output states, does not depend on their displacement~\cite{twamley,Jeong}. 
Thus we just perform the teleportation of a squeezed state with a particular displacement; $\alpha_0 =0$ in our experiment.

Furthermore the coordinates, namely the angle $\theta$, can be experimentally adjusted in such a way that the angle for the output is coincided with that for the input. 
Note that this adjustment of the angle is also made with the strong (classical) field. 
As a result, the fidelity does not also depend on the (relative) angle~\cite{twamley,Jeong}. 
Therefore we carry out the teleportation experiment with $\theta=0$.

Accordingly we just examine the teleportation of a squeezed thermal state with the variances of $\sigma_{\mathrm{in}}^{x}$ and $\sigma_{\mathrm{in}}^{p}$ of Eq.~(\ref{eq:sq}). 
In a broad sense, such states include the states whose squeezed variances are not smaller than the vacuum variance, namely $\sigma_{\mathrm{in}}^{x}\ge1/4$. In the present work, however, we teleport a squeezed vacuum state with the squeezed variance $\sigma_{\mathrm{in}}^{x}<1/4$ (see Fig.~2).

\section{TELEPORTATION PROCEDURE}

In this section, we briefly summarize the procedure of CV teleportation and describe our experimental set-up. 
The scheme of the procedure is outlined in Fig.~1. 
\begin{figure}[tbp]
\includegraphics[width=0.8\linewidth]{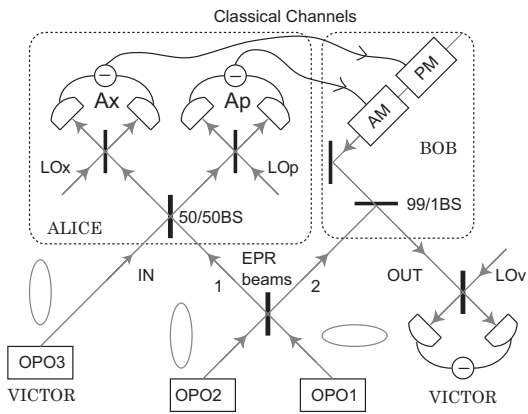}
\caption{\label{fig:1} Schematic setup of the experiment for quantum teleportation of a squeezed state. OPOs represent optical parametric oscillators. AM and PM denote amplitude and phase modulators. LOs are local oscillators for homodyne detection. The ellipses indicate the squeezed quadrature of each beam. All beam splitters except 99/1BS are 50/50 beam splitters. 
Symbols and abbreviations are defined in the text.}
\end{figure}
First sender Alice and receiver Bob share entangled EPR beams, which can be generated by combining two squeezed states at a 50/50 beam splitter (BS). 
One of the EPR beams is sent to Alice (mode 1) and the other is to Bob (mode 2). 
For the purpose of verifying the protocol, an input state is created by Victor (the ``verifier"). 
Note that an input state is unknown to both Alice and Bob in an ideal case. 
Alice combines the mode 1 and an input state at a 50/50 BS, and then measures $x$ and $p$ quadratures by two homodyne detectors; $x$ for one beam (Ax) and $p$ for the other (Ap). 
This measurement corresponds to the Bell-state measurement for CVs~\cite{braun98}. 
The output photocurrents of measurement results are transmitted to Bob through classical channels. 
After receiving the classical information from Alice, Bob reconstructs the teleported output state by performing a phase space displacement on the mode 2 beam. 
His displacement process consists of two parts. 
One is amplitude and phase modulations of a light beam~(AM and PM) based on the classical information from Alice, because we define the quantum state to be frequency sidebands at $\pm 1\mathrm{MHz}$ (with a bandwidth of 30kHz). 
The other is the coherent combination of this modulated beam and the mode 2 at a highly reflecting mirror (a 99/1 BS in our experiment).

In the absence of losses, the variances $\sigma_{\mathrm{out}}^{x}$ and $\sigma_{\mathrm{out}}^{p}$ associated with the output state are written by~\cite{furusawa98}
\begin{eqnarray}
\sigma_{\mathrm{out}}^{x} & = & g_x^2 \sigma_{\mathrm{in}}^{x} +\frac{e^{-2r_-}(1+g_x)^2}{8} + \frac{e^{+2r_+}(1-g_x)^2}{8}, \label{output-x} \\
\sigma_{\mathrm{out}}^{p} & = & g_p^2 \sigma_{\mathrm{in}}^{p} +\frac{e^{-2r_-}(1+g_p)^2}{8} + \frac{e^{+2r_+}(1-g_p)^2}{8}, \label{output-p}
\end{eqnarray}
where $r_{-},\ r_+$ are the squeezing and the antisqueezing parameters for squeezed states used to generate the EPR beams. 
$g_x$ and $g_p$ are suitably normalized gains of classical channels and defined as $g_x =\langle \hat{x}_{\mathrm{out}} \rangle /\langle \hat{x}_{\mathrm{in}} \rangle$ and $g_p =\langle \hat{p}_{\mathrm{out}} \rangle /\langle \hat{p}_{\mathrm{in}} \rangle$. 
When the gains are adjusted to unity, the displacement of an input state is properly reconstructed at Bob's station.

Finally Victor analyzes an output state from Bob's station. 
In our experiment, we verify that the squeezed variance of a squeezed vacuum state is properly transferred. 
When the state is teleported with $g_x =g_p =1$, the teleported state should show the smaller variance in $x$ quadrature than that for the case of a vacuum input ($r=0,\ \beta \to \infty$): 
\begin{equation}
(\sigma_{\mathrm{out}}^{x})_{sq} < (\sigma_{\mathrm{out}}^{x})_{vac}.
\label{relation}
\end{equation}
This is because the variance $\sigma_{\mathrm{in}}^{x}$ in the present work is smaller than the vacuum variance of 1/4. Although some losses are inevitable and the gains might slightly differ from unity in real experiments, the relation of Eq.~(\ref{relation}) should be satisfied in our work. 
Similarly, the inequality of $(\sigma_{\mathrm{out}}^{p})_{sq} > (\sigma_{\mathrm{out}}^{p})_{vac}$ should be expected to hold. 
We verify that the quantum teleportation process satisfies these inequalities.

We also verify success of quantum teleportation by using a fidelity. When an input state is a mixed state, the fidelity is described as follows~\cite{jozsa}:
\begin{equation}
F = \left\{ {\rm Tr} \left[ \left( \sqrt{\hat{\rho}_{\mathrm{in}}} \hat{\rho}_{\mathrm{out}} \sqrt{\hat{\rho}_{\mathrm{in}}} \right)^{1/2} \right] \right\}^2.
\label{fidelity1}
\end{equation}
This is an overlap between an input state $\hat{\rho}_{\mathrm{in}}$ and an output state $\hat{\rho}_{\mathrm{out}}$. If an input is a pure state $|\psi_{\mathrm{in}} \rangle$, the fidelity $F$ becomes $F=\langle \psi_{\mathrm{in}} | \hat{\rho}_{\mathrm{out}} | \psi_{\mathrm{in}} \rangle$. In the ideal quantum teleportation, the fidelity goes to unity, {\it F}=1, whereas {\it F}=0 means that the teleported state is orthogonal to the input state.

Our experimental setup is shown in Fig.~1. 
We generate three independent squeezed vacuum states. 
One of these states is used as an input for teleportation and the other two are used to produce entangled EPR beams. 
In order to generate each squeezed vacuum state, we use a subthreshold degenerate optical parametric oscillator (OPO) with a 10mm long potassium niobate crystal ($\mathrm{KNbO_3}$). 
The crystal is temperature-tuned for type-I noncritical phase matching. 
Each OPO cavity is a bow-tie-type ring cavity which consists of two spherical mirrors (radius of curvature 50mm) and two flat mirrors. 
The round trip length is 500mm and the waist size in the crystal is 20$\mu$m. 
The output of a Ti:Sapphire laser at 860nm is frequency-doubled in an external cavity with the same configuration as the OPOs and divided into three beams to pump the three OPOs. 

EPR beams are generated by combining two squeezed vacuum states from OPO1 and OPO2 at a 50/50 BS with a $\pi$/2 phase shift as shown in Fig.~1. We characterize the quantum entanglement with the inseparability criterion proposed in Refs.~\cite{duan00,simon00} and obtain the result of $\langle [ \Delta (\hat{x}_1 -\hat{x}_2 ) ]^2 \rangle+\langle [ \Delta (\hat{p}_1 +\hat{p}_2 ) ]^2 \rangle=0.47\pm 0.04<1$. 
This result shows the existence of quantum entanglement between the EPR beams.

The normalized gains of two classical channels are adjusted in the manner of Ref.~\cite{zhang03}. 
We obtain the measured gains of $g_x =0.98 \pm0.04$ and $g_p =0.98 \pm0.03$, respectively. 
For simplicity, these gains are fixed throughout the experiment and treated as unity.

\section{RESULTS}

Before performing teleportation, we first measure the input squeezed vacuum state at the Ap homodyne detector at Alice by removing Alice's 50/50 BS. 
Figure 2 shows the measurement results. 
\begin{figure}[t]
\includegraphics[width=0.9\linewidth]{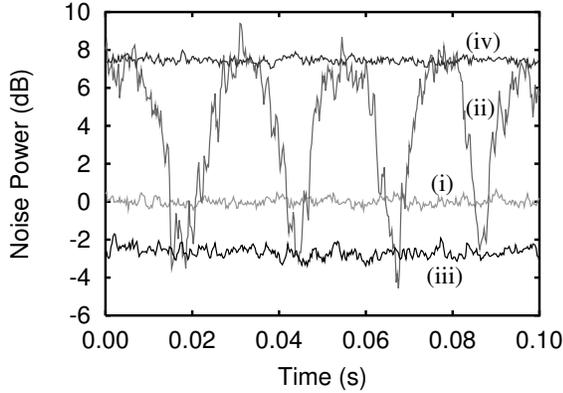}
\caption{\label{fig:2} The measurement results on the input squeezed state recorded by the Ap homodyne detector with Alice's 50/50 BS removed. Trace (i) shows the corresponding vacuum noise level; trace (ii) is the variance of the squeezed state with the LO phase scanned; traces (iii) and (iv) are the minimum and the maximum noise levels with the LO phase locked. The measurement frequency is 1 MHz, and the resolution and video bandwidth are 30 kHz and 300 Hz, respectively. All traces except for (ii) are averaged ten times.}
\end{figure}
The squeezing and antisqueezing are $-2.66\pm$0.49dB and 7.45$\pm$0.17dB, respectively, compared to the vacuum noise level. 
Visibility between the input and LO at Ap detector is 0.968. 
From this visibility and the results, we infer $-2.92\pm 0.56$dB of squeezing $\sigma_{\mathrm{in}}^{x}$ and 7.68$\pm$0.27dB of antisqueezing $\sigma_{\mathrm{in}}^{p}$, where we assume that $x$ quadrature is squeezed. 
In the teleportation experiment, the relative phase of the input is properly adjusted and locked.

We then proceed to the experiment of teleportation. 
Figure 3 shows the Victor's measurement results of the output state from Bob. 
First we consider the case without EPR beams in Fig. 3(a), which is so-called classical teleportation. 
In this case, the variances are somewhat large due to the quduty that must be paid for crossing the border between quantum and classical domains~\cite{braun98}. 
We observe the noise levels of 4.86$\pm$0.20dB and 4.92$\pm$0.20dB in $x$ and $p$ quadratures for the vacuum input. 
These levels correspond to three units of the vacuum variance~(4.77dB in a logarithmic scale): one from the input intrinsic variance and two units from the quduty~\cite{braun98}. 
For the squeezed vacuum input, we observe the minimum noise level of 4.12$\pm$0.23dB in $x$ quadrature and the maximum noise level of 8.92$\pm$0.16dB in $p$ quadrature in this classical teleportation.

Next we carry out quantum teleportation with EPR beams as shown in Fig. 3(b). 
In the case of a vacuum input, the output variances are reduced from the classical case due to the quantum entanglement. 
We obtain the noise levels of 2.90$\pm$0.21dB and 3.01$\pm$0.19dB, respectively, for $(\sigma_{\mathrm{out}}^{x})_{vac}$ and $(\sigma_{\mathrm{out}}^{p})_{vac}$. 
The noise reduction from the classical teleportation indicates the success of quantum teleportation of the vacuum. 
Based on these variances and the measurement results $\sigma_{\mathrm{in}}^{x}$,\ $\sigma_{\mathrm{in}}^{p}$ of the squeezed input, we can calculate the expected output variances $(\sigma_{\mathrm{out}}^{x})_{sq}$ and $(\sigma_{\mathrm{out}}^{p})_{sq}$ for our squeezed input using Eqs.~(\ref{output-x})~and~(\ref{output-p}). 
The calculated variances are 1.71$\pm0.58$dB and 8.24$\pm0.31$dB for $(\sigma_{\mathrm{out}}^{x})_{sq}$ and $(\sigma_{\mathrm{out}}^{p})_{sq}$, respectively. 
\begin{figure}[t]
\begin{center}
\includegraphics[width=0.90\linewidth]{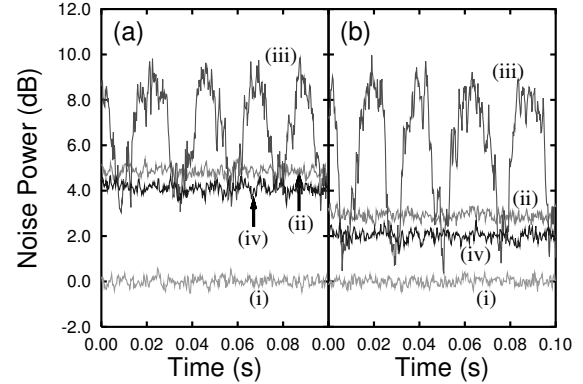}
\caption{\label{fig:3} The measurement results on the output states recorded by Victor in $x$ quadrature, where (a) classical teleportation without the EPR beams, (b) the teleportation with the EPR beams; (i) the corresponding shot noise level, (ii) the vacuum state input, (iii) the squeezed state input with the phase of the input state scanned, and (iv) the minimum noise levels with the phase of the input state locked. The measurement conditions are the same as for Fig. 2.}
\end{center}
\end{figure}

The squeezed vacuum state shown in Fig. 2 is subsequently teleported. 
We obtain the minimum noise level of 2.03$\pm$0.24dB for $(\sigma_{\mathrm{out}}^{x})_{sq}$ and the maximum noise level of 8.18$\pm$0.17dB for $(\sigma_{\mathrm{out}}^{p})_{sq}$ (not shown), respectively, which are in good agreement with the expected variances. 
The squeezed variance of the teleported state is clearly smaller than that of the teleported vacuum state in $x$ quadrature, and then the inequality $(\sigma_{\mathrm{out}}^{x})_{sq}<(\sigma_{\mathrm{out}}^{x})_{vac}$ is satisfied. 
Similarly the inequality $(\sigma_{\mathrm{out}}^{p})_{sq}>(\sigma_{\mathrm{out}}^{p})_{vac}$ holds. 
Therefore the squeezed variance of our squeezed vacuum input is certainly teleported. 
We verify that the teleportation process operates properly for the nonclassical input of the squeezed vacuum.

Note that the observed squeezed variance of the teleported state is larger than Victor's shot noise level, which shows that the output state is not a nonclassical state. 
In order to get a nonclassical state at Bob's place, the quantum entanglement with stronger nonclassical correlation is required. 
If an input squeezed state and two squeezed states used for the generation of EPR beams have the same degree of squeezing, more than 4.77dB squeezing is needed. 
It is a next challenge to generate a teleported state whose variance is below the shot noise level.

We next evaluate the success of the teleportation process by a fidelity. The process can be regarded as a generalized thermalizing quantum channel~\cite{ban02} and does not alter the Gaussian character of an input state. Assuming both input and output states show Gaussian distribution without displacement, we can characterize these states by measuring the variances $\sigma^x_j$ and $\sigma^p_j~(j=\mathrm{in},\ \mathrm{out}$) like Eq.~(\ref{eq:sq}).

Since a vacuum is one of coherent states, the fidelity $F_{vac}$ for the vacuum teleportation at unity gains is simply given by~\cite{braun01}
\begin{equation}
F_{vac} =\frac{2}{\sqrt{(1+4\sigma_{\mathrm{out}}^{x})(1+4\sigma_{\mathrm{out}}^{p})}}. 
\end{equation}
From the measured variances, we obtain the fidelity of $0.67\pm 0.02$ for quantum teleportation of a vacuum input, which exceeds the classical limit of 0.5~\cite{braun00,braun01}. 
This result clearly shows the success of quantum teleportation of a vacuum state. 
But we cannot apply this classical limit to the case of the squeezed state input, since the fidelity depends on an input state like Eq.~(\ref{fidelity1}).

It could be considered that a (mixed) squeezed vacuum input is transformed into a squeezed thermal state through the imperfect teleportation process. 
For these squeezed states, the fidelity $F_{sq}$ in Eq.~(\ref{fidelity1}) can be calculated explicitly as follows~\cite{twamley}:
\begin{eqnarray}
F_{sq} & = & \frac{2 \sinh (\beta_{\mathrm{in}}/2) \sinh (\beta_{\mathrm{out}}/2)}{\sqrt{Y}-1}, \label{fidelity2} \\
Y &=& \cosh^2 (r_{\mathrm{in}} -r_{\mathrm{out}} ) \cosh^2 [(\beta_{\mathrm{in}} +\beta_{\mathrm{out}}) /2] \nonumber \\
& & {}- \sinh^2 (r_{\mathrm{in}} -r_{\mathrm{out}} ) \cosh^2 [(\beta_{\mathrm{in}} -\beta_{\mathrm{out}})/2]. \nonumber
\end{eqnarray}
From Eq.~(\ref{eq:sq}), the squeezing parameter $r_j$ and the inverse temperature $\beta_j$ are given by
\begin{equation}
r_j=\frac{1}{4} \ln \Bigl( \frac{\sigma_{j}^{p}}{\sigma_{j}^{x}} \Bigr),\ \beta_j = \ln \Bigl( 1+\frac{2}{4\sqrt{\sigma_{j}^{x} \sigma_{j}^{p}}-1} \Bigr), 
\end{equation}
where $j=(\mathrm{in},\ \mathrm{out})$. 
Thus we can calculate the fidelity from the measured variances.

We apply the fidelity $F_{sq}$ to the particular input state in this experiment. The fidelity for the ``perfect'' classical teleportation could be calculated from the measurement results on the input state shown in Fig.~2. 
The fidelity for the classical teleportation $F_{sq}^C$ is calculated as $0.73\pm 0.04$, and we regard this value as the classical limit for the input. 
The measured fidelity of the classical case without the EPR beams is $0.73\pm 0.05$, which is in good agreement with the classical limit. 
In the quantum case with the EPR beams, we obtain the result of $F_{sq}^Q=0.85\pm 0.05$ which is higher than the classical limit. 
This fact shows the success of quantum teleportation and it means that the teleported state in the quantum teleportation is more close to the input state than that in the classical teleportation.

\section{DISCUSSION}

Here we discuss the classical limit for a set of squeezed thermal states. 
The classical limit for the set has not been investigated so far. 

In the case of a set of coherent states, the classical limit of 0.5 is derived by averaging the fidelity for coherent states randomly chosen from the set~\cite{braun00,braun01}. 
A coherent state is characterized only by one parameter, i.e., its displacement. 
Thus the displacement is randomly selected in the derivation of the limit, and the fidelity of each state is averaged over the set. 

From Eq.~(\ref{eq:wigner}), however, a squeezed thermal state is fully characterized by four parameters; displacement~$\alpha_0$, an angle~$\theta$, inverse temperature~$\beta$ and squeezing~$r$. 
In order to derive the classical limit for a set of squeezed thermal states, these four parameters should be chosen randomly or with some probability, and the classical limit is calculated by averaging the fidelity over the set. 
However, as mentioned in Sec. II, the fidelity does not depend on $\alpha_0$ and $\theta$ at the appropriate setting in the teleportation process. 
Thus we just consider only two parameters $\beta$ and $r$. 

Now we investigate the dependence of the fidelity $F_{sq}$ on the variance $\coth(\beta/2)$ and the antisqueezing $e^{+2r}$ rather than $\beta$ and $r$. 
In particular, $\coth(\beta/2)$ is the variance of a thermal state, normalized to the vacuum variance, and it may indicate the mixedness of a squeezed thermal state. 
From Eq.~(\ref{eq:sq}) and the results of Fig.~2, our input squeezed vacuum shows 2.39$\pm$0.31dB of $\coth(\beta/2)$ and 5.06$\pm$0.26dB of $e^{+2r}$.

Using these values and Eqs.~(\ref{output-x}) and~(\ref{output-p}), we can calculate the fidelity $F_{sq}^C$ for ``perfect'' classical teleportation, which is plotted as a function of $\coth(\beta/2)$ and $e^{+2r}$ in Fig.~4. 
\begin{figure}[t]
\includegraphics[width=0.85\linewidth]{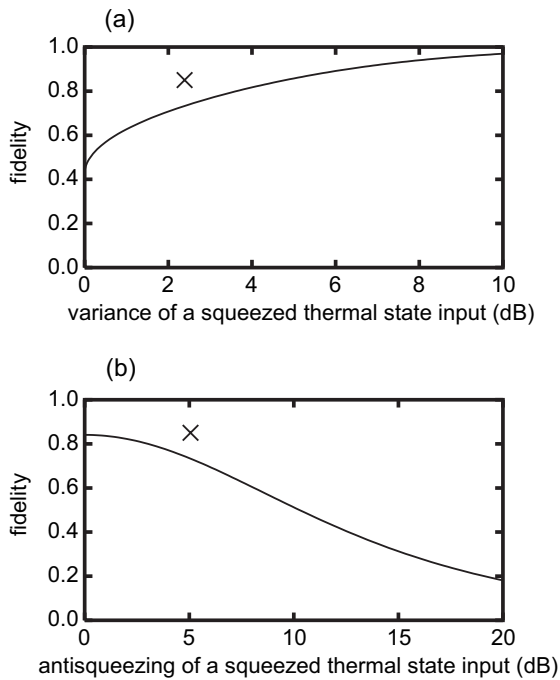}
\caption{\label{fig:4} The calculated fidelity $F_{sq}^C$ between the input and the output. 
(a) the dependence of $F_{sq}^C$ on the variance $\coth (\beta/2)$ with the fixed antisqueezing $e^{+2r}$ of 5.06dB. 
The measured value of the fidelity $F_{sq}^Q$ is plotted as the cross ``$\times$'' at $\coth (\beta/2)$ of 2.39dB. 
(b) the dependence of $F_{sq}^C$ on the antisqueezing $e^{+2r}$ with the fixed variance $\coth (\beta/2)$ of 2.39dB. 
The measured value of the fidelity $F_{sq}^Q$ is plotted as the cross ``$\times$'' at $e^{+2r}$ of 5.06dB. 
}
\end{figure}
The trace in Fig.~4(a) shows the dependence of $F_{sq}^C$ on the variance $\coth (\beta/2)$ with the fixed antisqueezing $e^{+2r}$ of 5.06dB. 
On the other hand, the trace in Fig.~4(b) shows the dependence of $F_{sq}^C$ on the antisqueezing $e^{+2r}$ with the fixed variance $\coth (\beta/2)$ of 2.39dB. 
The measured fidelity $F_{sq}^Q =0.85\pm 0.05$ is plotted as the cross ``$\times$'', which is larger than the fidelity $F_{sq}^C =0.73\pm 0.04$ for our input state.

In Fig.~4(a), the fidelity $F_{sq}^C$ increases from 0.44 at 0dB of $\coth (\beta/2)$, as the variance $\coth (\beta/2)$ or the mixedness increases. 
The variance of 0dB corresponds to a pure squeezed vacuum with 5.06dB squeezing. 
The fidelity of 0.44 at 0dB is slightly smaller than 0.5 for a vacuum state, which indicates that a pure squeezed state is more fragile than a vacuum state through a classical teleportation process. 
This agrees with the fact that it is not possible to transfer the nonclassical feature of squeezing in classical teleportation. 
In the limit of $\coth (\beta/2) \to \infty$ or $T \to \infty$ where an input state may become completely mixed and classical, the fidelity $F_{sq}^C$ goes to unity. 
Note that the fidelity is very sensitive to small amounts of mixedness~\cite{Jeong}.

On the other hand, in Fig.~4(b), the fidelity $F_{sq}^C$ decreases from 0.84 at 0dB of $e^{+2r}$, as the antisqueezing increases. 
The tendency of $F_{sq}^C$ means that it is more difficult to transfer a squeezed thermal state with a larger degree of squeezing through the classical teleportation. 
The value of 0.84 at 0dB of $e^{+2r}$ is much larger than 0.5, which shows that a thermal state can be more easily teleported than a vacuum through a classical teleportation process. 
This fact indicates that the fidelity varies sensitively with the mixedness~(2.39dB of $\coth (\beta/2)$ in our experiment).

As mentioned before the classical limit can be obtained by averaging the fidelity $F_{sq}^C$ over the set of input states which are characterized by $\beta$ and $r$, or the variance $\coth(\beta/2)$ and the antisqueezing $e^{+2r}$. 
The question is to what extent the set should covers values of $\coth(\beta/2)$ and $e^{+2r}$. 
Concerning the variance of $\coth(\beta/2)$, some states with $\coth(\beta/2) \gg 1$ would be treated as classical thermal states. 
They could be easily teleported in classical teleportation with almost unity fidelity as shown in Fig.~4(a). 
If the set of interest contains such classical states, the value of the classical limit would increase. 
On the other hand, the fidelity shows a decreasing tendency with the antisqueezing $e^{+2r}$ as shown in Fig.~4(b). When the fidelity is averaged over the whole range of $e^{+2r}$, the classical limit would become small value. 
Therefore the classical limit depends strongly on an input set and the situation where the set is used. 
It is difficult to calculate the general classical limit for an input set which includes the state with $\coth(\beta/2)$ and $e^{+2r}$ chosen from the whole range of them.

Finally note that the fidelity is strongly dependent on a set of input states, and very sensitive to mixedness of the states. 
Therefore the fidelity may not be good measure of a success criterion of quantum teleportation, especially for a mixed state, as argued in Ref.~\cite{Jeong}. 
The establishment of the criterion for the mixed state is needed.

\section{CONCLUSION}

We have demonstrated CV quantum teleportation of a squeezed vacuum state and calculated the fidelity for the state. The measured fidelity of the squeezed state input in the quantum teleportation is $0.85\pm 0.05$ which is higher than the classical case $0.73\pm 0.04$. This result shows that the teleported state in the quantum teleportation process is more similar to the input state than that in the classical teleportation. 

We have also established an operational method of evaluation for quantum teleportation of a squeezed thermal state, and discussed the classical limit for a set of squeezed thermal states. 
Four parameters defined in Eqs.~(\ref{eq:sq}) and (\ref{eq:wigner}) can be experimentally tunable. 
For example, the antisqueezing $e^{+2r}$ can be varied with pump power of OPO, and displacement $\alpha_0$ can be adjusted through a displacement operation. 
However the general classical limit and the success criterion~(with or without fidelity) for mixed states remain a topic for future study.

Furthermore we have observed the smaller variance of a teleported squeezed state than that for the case of a vacuum state input. This means that the nonclassical feature of a squeezed vacuum state is preserved throughout the teleportation process. 
It is a next challenge to generate a teleported state whose variance is below the shot noise level. 
In CV quantum teleportation and quantum information processing, it is important to transfer the squeezing and to reconstruct a squeezed state at the receiving station. 

\begin{acknowledgments}
This work was partly supported by the MEXT and the MPHPT of Japan, and Research Foundation for Opto-Science and Technology. We would like to thank D. E. Browne, M. Murao, S. L. Braunstein and M. Sasaki for valuable comments and discussions. 
\end{acknowledgments}

\end{document}